\documentclass[prb,preprint,doublespace,superscriptaddress,a4paper,showpacs]{revtex4}
\usepackage{amsmath,graphicx,setspace}
\usepackage{color}
\begin{document}
%\draft
%
\title{The Magneto-coulomb effect in spin valve devices}

\author{S.J. van der Molen}
\affiliation{Physics of Nanodevices, Materials Science Centre,
Rijksuniversiteit Groningen, Nijenborgh 4, 9747 AG Groningen, The
Netherlands}
\author{N. Tombros} \affiliation{Physics of
Nanodevices, Materials Science Centre, Rijksuniversiteit
Groningen, Nijenborgh 4, 9747 AG Groningen, The Netherlands}
\author{B.J. van Wees}
\affiliation{Physics of Nanodevices, Materials Science Centre,
Rijksuniversiteit Groningen, Nijenborgh 4, 9747 AG Groningen, The
Netherlands}
\date{\today}
\begin{abstract}
We discuss the influence of the magneto-coulomb effect (MCE) on
the magnetoconductance of spin valve devices. We show that MCE can
induce magnetoconductances of several per cents or more, dependent
on the strength of the coulomb blockade. Furthermore, the
MCE-induced magnetoconductance changes sign as a function of gate
voltage. We emphasize the importance of separating conductance
changes induced by MCE from those due to spin accumulation in spin
valve devices.
\end{abstract}
 \maketitle
\begin{figure}[h]
\begin{center}
\includegraphics[width=8cm]{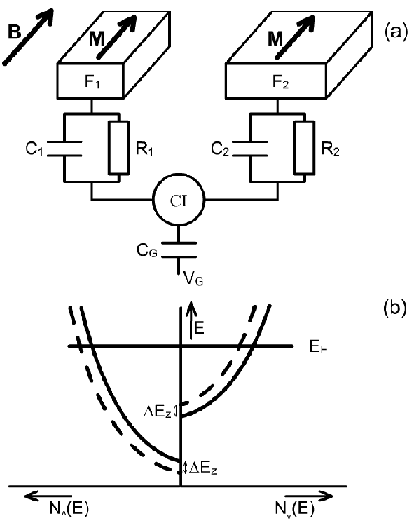}
\end{center}
\caption{a) The sample structure considered. Two ferromagnetic
strips, $F_1$ and $F_2$, with coercive fields $B_{c1}$ and
$B_{c2}$ are weakly connected to a coulomb island (CI) via two
tunnel barriers (resistances $R_1$ and $R_2$ and capacitances
$C_1$ and $C_2$). Furthermore, a gate connects capacitively to the
island ($C_G$). b) Sketch of the density of states N of the two
spin species in a ferromagnet, versus energy. When a magnetic
field is applied, the energies of the two spin species shift
($\Delta E_z$) in opposite directions by the Zeeman effect. Since
$N^\uparrow > N^\downarrow$, this results in a change in the work
function, $\Delta W$.} \label{mcevsb}\end{figure}

The recent past has seen an impressive effort in connecting
ferromagnetic leads to ever smaller non-ferromagnetic structures.
The main idea behind this is to make use of the electron spin for
device purposes. In a two-terminal, spin valve geometry, a
resistance difference $\Delta R$ is expected between two basic
situations. First, if the two ferromagnetic leads are magnetized
in an anti-parallel fashion, the majority spin species injected at
the first ferromagnet is predominantly reflected at the second
ferromagnet. This results in a high resistance state. On the other
hand, in the case of parallel magnetizations, the injected
majority spin couples well to the second ferromagnet, leading to a
lower resistance state. With the miniaturization of the central
structure, quantum confinement effects come into play. Recently,
quite some progress has been made in studying spin devices in the
presence of coulomb blockade.
\cite{tsukagoshinat,tsukagoshiapl,sahooapl,sahoonat,yakushiji,zhang,
kim,chakraborty,orgassa,zhao,alphenaar} The interpretation of the
two-terminal data in these reports has mostly focused on spin
transport and spin accumulation. Here, we discuss another
influence on the two-terminal resistance in ferromagnetically
contacted nanostructures, namely the magneto-coulomb effect (MCE)
discovered by Ono \emph{et al.}\cite{ono}.
\\
In this contribution, we consider a confined conductor weakly connected to
two ferromagnets, $F_1$ and $F_2$ (see Fig. \ref{mcevsb}a). The
coupling is described by two sets of resistances and capacitances,
$R_1$, $C_1$ and $R_2$, $C_2$, respectively. Furthermore, the island
can be gated by a voltage $V_g$ via a capacitor $C_g$. For a basic
introduction to the MCE, we first concentrate on one of the
ferromagnets only, $F_1$, which is assumed magnetized in the
positive direction. Let us suppose that a positive external magnetic
field ($B>0$) is applied. In that case, the energy of the spin
up($\uparrow$) and spin down($\downarrow$) electrons shift by the
Zeeman energy, in opposite directions (see Fig. \ref{mcevsb}b).
However, for a ferromagnet, the density of states of both spin
species differs ($N^\uparrow>N^\downarrow$). Hence, a shift in the
chemical potential $\Delta \mu$ needs to take place to keep the
number of electrons constant:\cite{ono}
\begin{equation} \label{eq:1}
\Delta \mu = - \frac{1}{2}P g \mu_B B
\end{equation}
where the thermodynamic polarization P is defined as $P =
\frac{N^\uparrow-N^\downarrow}{N^\uparrow+N^\downarrow}$,\cite{voetP}
$g$ is the gyromagnetic ratio and $\mu_B$ is the Bohr magneton. In
practice, however, the ferromagnet will be attached to a
macroscopic non-magnetic lead. This demands equal
chemical potentials in both metals. Hence, the energy shift in the
ferromagnet translates to a change in the contact potential between
the ferromagnet and the normal metal, $\Delta \phi$, according to,
$-e\Delta \phi=-\Delta \mu$.\cite{ono} Equivalently, one could say
that the work function of the ferromagnet changes by $\Delta
W=-\Delta \mu$. Since the ferromagnet is weakly coupled to the
central island, this shift influences the Coulomb levels of the
latter. In fact, an additional charge $\Delta q$ is induced onto the island
due to the contact potential change $\Delta \phi$. Applying a magnetic field
thus has an effect that is similar to changing the gate voltage. This equivalence
has been beautifully demonstrated by Ono \emph{et al.}\cite{ono} For the situation
 sketched above, we find:
\begin{equation} \label{eq:2}
\Delta q(B) = \frac{C_1}{2e} P g \mu_B B
\end{equation}

\begin{figure}[h]

\begin{center}
\includegraphics[width=8cm]{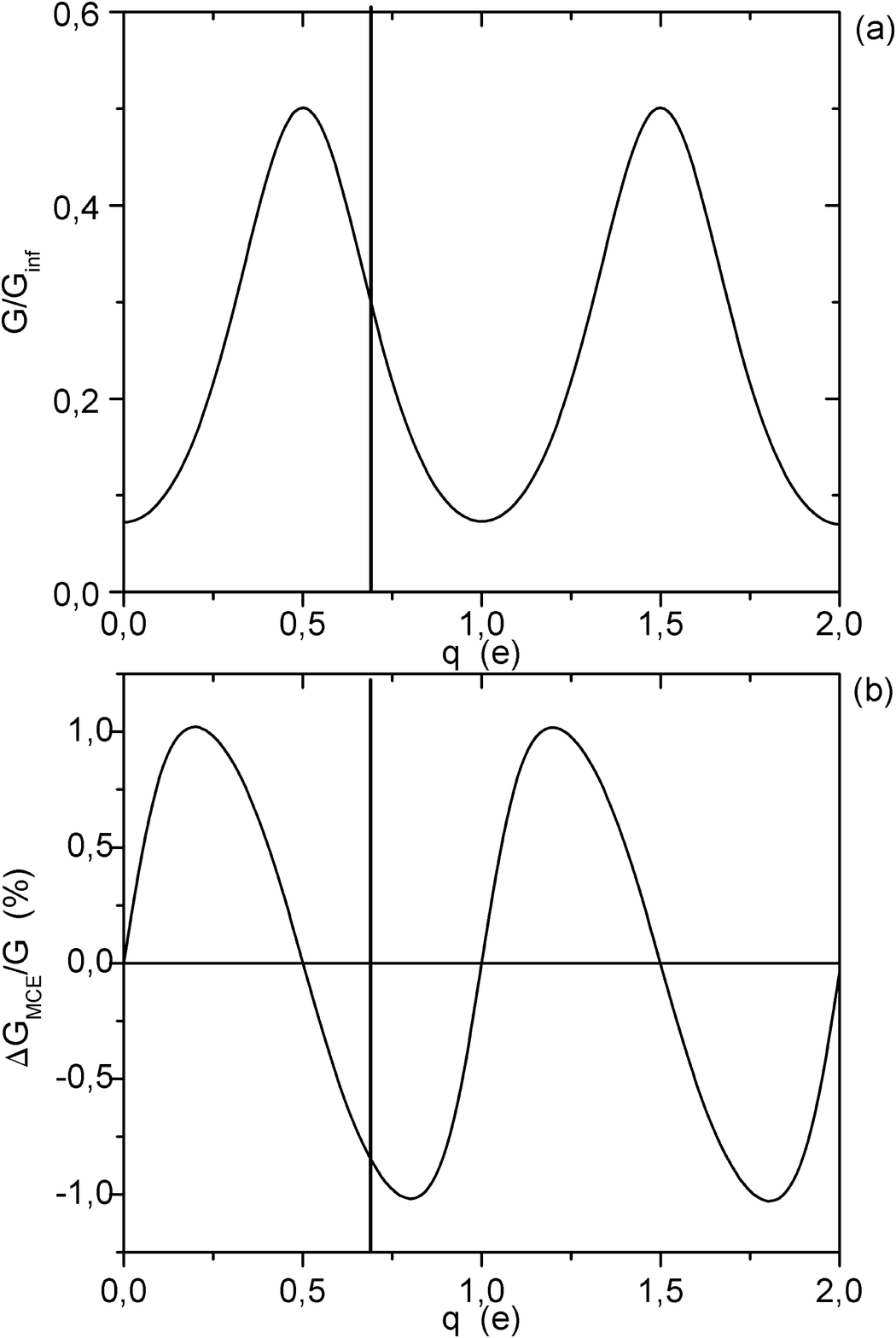}
\end{center}
\caption{a) G vs charge state q calculated for the system in Fig.
\ref{mcevsb}a) with parameters: $C_1=C_2=2 \cdot 10^{-17} F, C_g=
5 \cdot 10^{-18} F, R_1 = R_2 = 2.5 M\Omega$. G is given in units
of $G_{\infty}=1/(R_1+R_2)=0.2 \mu S$ b) Relative conductance
change $\Delta G_{MCE} /G$ vs $\Delta E_F$ (in $\%$). We use
$P=-0.6$, $B_{c1}=0.09 T$, $B_{c2}=0.11 T$ and take $g=2$. Note
that the relative resistance change, $\Delta R_{MCE}/R$, equals
$-\Delta G_{MCE} /G$. Figure \ref{vsB} is evaluated at $q=0.69e$
$e^2/C_{tot}$, indicated by the vertical lines in a) and
b).}\label{tmr}
\end{figure}

\begin{figure}[h]

\begin{center}
\includegraphics[width=8cm]{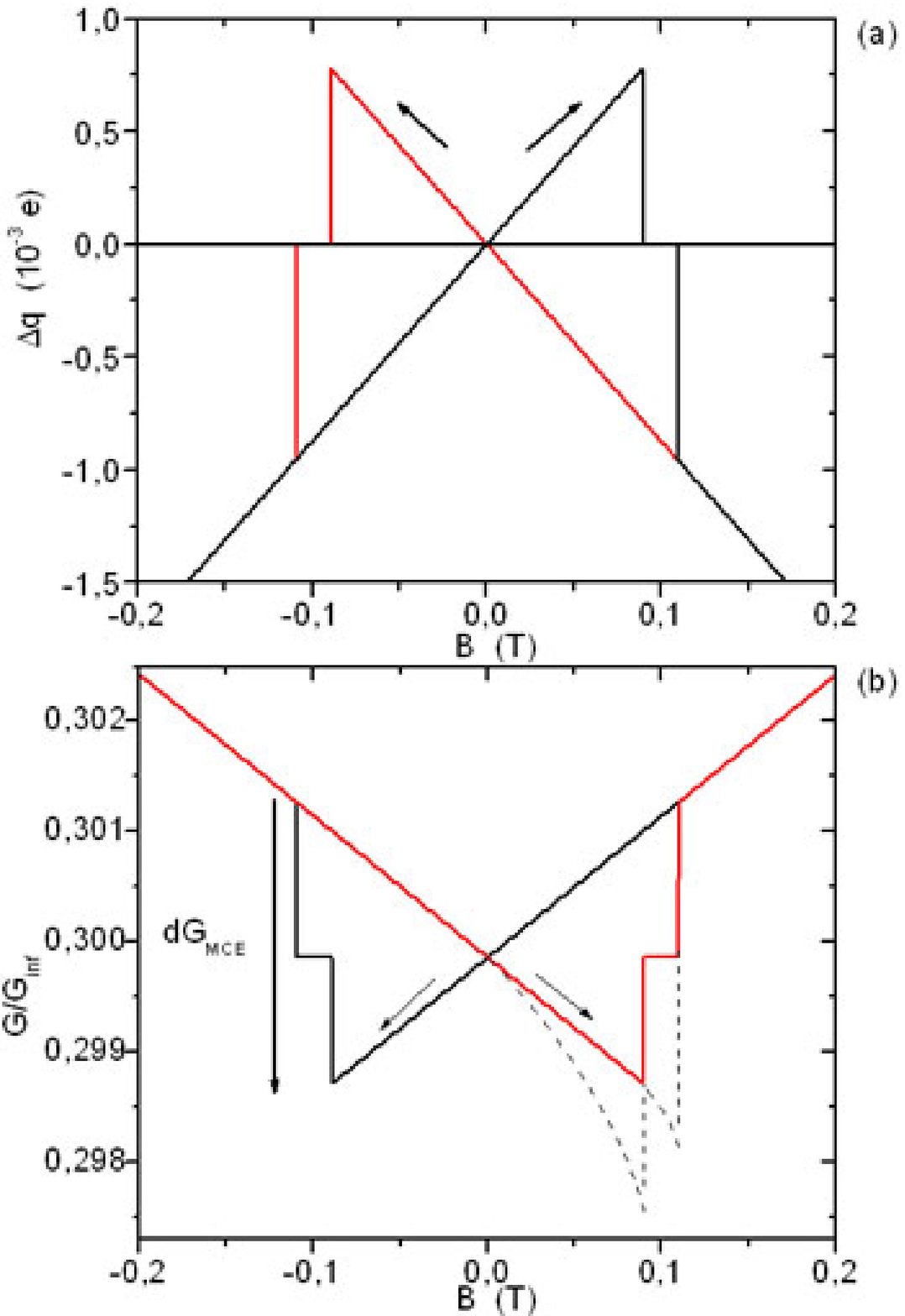}
\end{center}
\caption{(Color online) a) Induced charge on the central island,
$\Delta q$, versus the applied magnetic field (see eq.\ref{eq:2}).
$\Delta q$ varies linearly with B, except at the switching fields,
where steps are seen. The curve ignores the demagnetizing field.
b) Influence of the external magnetic field on the zero-bias
conductance, calculated with eq. \ref{eq:3}. Solid line:
demagnetization field ignored. We define the sum of these steps as
$\Delta G_{MCE}<0$. We note that $|\Delta G_{MCE}|$ can become
quite large, despite the low values of the induced charge, since
it depends critically on the sharpness of the Coulomb peaks (see
Fig. \ref{tmr}). Dashed line: qualitative effect of the rotation
of the demagnetization field at the nanotube (only drawn for
positive fields). Graph a) and b) are evaluated at $q=0.69 e$,
indicated in Fig. \ref{tmr}} \label{vsB}
\end{figure}

Hence, if no magnetization rotation or switching takes place in
the ferromagnet, the induced charge onto the island
changes linearly with the applied field B. Interestingly, for a
system in the coulomb blockade regime, the conductance G(q) is a
(more or less periodic) function of the induced
charge. Combining G(q) with eq. \ref{eq:2}, we find that the conductance changes with field:
\begin{equation} \label{eq:3}
\Delta G(B) = \frac{dG}{dq} \Delta q(B)
\end{equation}
For a Coulomb island, $G(q)$ can be calculated (or it can be
measured experimentally versus the gate voltage). The exact theory
to apply depends on the magnitude of the various energy scales
involved.\cite{kouwenhoven} In any case, the sign of $\Delta G$ is
determined by the signs of both P and $\frac{dG}{dq}$. Since the
function $G(q)$ is periodic, $\frac{dG}{dq}$ and $\Delta G$ change
sign periodically, specifically at a Coulomb peak.\\
\\
Next, we incorporate magnetization switching. Again, we start with
ferromagnet $F_1$ magnetized in the positive direction, but now we
ramp down the external field ($B<0$). Then, according to eq.
\ref{eq:3}, the conductance changes linearly with B, as long as
the magnetization of the ferromagnet is unchanged. However, when B
reaches the coercive field, i.e., $B=-B_c$,  the magnetization of
the ferromagnet switches to the negative direction. Hence, also
$\Delta q$ changes discontinuously, by $\Delta q_c= \frac{C_1}{e}
Pg\mu_B B_c$. This results in a jump in the conductance via eq.
\ref{eq:3}. For more negative B fields, the conductance change
will be linear with B again, but now with opposite sign. So far,
we have considered an island connected to one ferromagnet only.
The extension to a spin valve device with two ferromagnetic
contacts is rather trivial, since their effects can be added.
Hence, a conductance change linear in B is expected,
with discontinuities at the coercive fields of both ferromagnets.\\
To illustrate the above, we consider the device in Fig.
\ref{mcevsb}a), where $F_1$ and $F_2$ have different switching
fields. (This can be achieved by choosing thin strips of different
widths).\cite{johnson,jedema,tombros} To calculate the conductance
properties of the system, i.e. $G(q)$, we make use of the orthodox
model of coulomb blockade.\cite{kouwenhoven,voetCB} This choice is
rather arbitrary, since eq. \ref{eq:3} can in principle be applied
to other regimes of coulomb blockade. In Fig. \ref{tmr}a), we show
G vs $q$  for a certain choice of (symmetric) system parameters
(see caption Fig. \ref{tmr}).\cite{voetq} From Fig. \ref{tmr}a)
and eq. \ref{eq:3}, we infer that the sign and magnitude of the
MCE depend critically on two properties: i) the system parameters,
which define the sharpness of the Coulomb peaks; ii) the charge
state about which $\Delta q$ applies, which defines the distance
to a Coulomb peak. Close to the inflection point of sharp Coulomb
peaks, $\frac{dG}{dq}$ can become very large. Therefore, even a
small $\Delta q$ can induce a sizeable resistance change,
\emph{without} a fundamental
limitation. In principle, effects exceeding $100 \%$ are possible.\\
Next, we determine the field dependence of the conductance in the
system considered. We (arbitrarily) evaluate around the
charge state $q = 0.69 e$, where $dG/dq <0$ (indicated in Fig.
\ref{tmr}). Furthermore, we use $P=-0.6$, which is the
thermodynamic polarization of cobalt.\cite{ono,voetP} In Fig.
\ref{vsB}a), we plot the induced charge on the island as a
function of magnetic field. Using eq. \ref{eq:3} together with Figs. \ref{tmr}a) and \ref{vsB}a) we obtain the field dependence of the conductance (see Fig. \ref{vsB}b). As discussed above, MCE gives
linear conductance changes for fields exceeding the switching
fields (giving a 'background magnetoconductance' for large
fields). Around the switching fields, however, discontinuous
changes are seen which lead to hysteretic behavior. We note that
Fig. \ref{vsB}b) does show similarities with several experiments
in spin valve devices\cite{voetoffset}. This emphasizes the
importance to separate both phenomena.\cite{tombros} To connect to
experiment, we define the conductance change due to MCE, $\Delta
G_{MCE}$, as the sum of the two conductance steps at the coercive
fields, i.e., $\Delta G_{MCE}=-\frac{dG}{dq} Pg\mu_B (C_1
B_{c1}+C_2 B_{c2})/e$. We indicate $\Delta G_{MCE}$ in Fig.
\ref{vsB}b). With this definition, we are able to plot the
relative magnetoconductance change $\Delta G_{MCE}/G$ as a
function of the charge state (see Fig. \ref{tmr}b)). Since this quantity is
proportional to the logarithmic derivative of the function $G(q)$,
it changes sign at the extremes of Fig. \ref{tmr}a).\cite{voetoffset} Figures \ref{tmr} and \ref{vsB} summarize the magnetoresistances that can be expected in
two-terminal spin valve structures, as a result of MCE.\\
\\
Recently, much work has been done to investigate magnetic field
induced conductance changes in quantum dot-like structures, such
as carbon nanotubes\cite{
tsukagoshinat,tsukagoshiapl,sahooapl,sahoonat,
kim,chakraborty,orgassa,zhao,alphenaar,morpurgo} and small metal
islands.\cite{yakushiji,zhang}. In these studies, conductance
changes are seen, which are generally interpreted in terms of spin
accumulation. However, three phenomena are noteworthy: 1) in many
cases, the change in conductance sets in before the magnetic field
changes sign, i.e. before the ferromagnetic electrodes switch
their magnetization. \cite{tsukagoshiapl,
tsukagoshinat,sahooapl,sahoonat,nagabhirava,yakushiji}. 2) In some
studies the magnetoconductance changes sign as a function of gate
voltage.\cite{tsukagoshiapl,sahoonat,nagabhirava,morpurgo} 3) In
carbon nanotubes connected to only one ferromagnet (and to gold),
field-induced conductance changes are also observed.\cite{jensen}
In the
latter system spin detection is clearly not possible.\\
We believe that in many experiments, MCE plays an important role.
As seen in Fig. \ref{tmr}, MCE-induced conductance changes have
the following properties: 1) they set in continuously at zero
field; 2) they change sign as a function of gate voltage, exactly
at the Coulomb peaks; 3) MCE-induced conductance changes also take
place for coulomb islands connected to only one ferromagnet, as
discussed above. Hence, the combination of MCE with spin
accumulation could be responsible for part of the phenomena listed
above. We note that the sign changes seen in
Refs.\cite{tsukagoshiapl,sahoonat,nagabhirava,morpurgo} have been
explained within (coherent) spin transport models (see also Ref. \cite{cottet}). However, in most of these systems coulomb blockade was also observed. This implies that MCE should be taken into account to obtain full correspondence between experiment and theory.\\
More generally, it is important to separate spin accumulation and MCE (and
other magnetoresistances) experimentally. The best way to do this, is by a direct measurement, using a non-local, four-probe geometry\cite{tombros}.
This method separates out all magnetoresistances, not only MCE. If a
non-local measurement is not possible, the MCE and spin accumulation
should be separated in other ways. For example by monitoring the
temperature and gate voltage dependence of the relative conductance
changes and comparing these data sets to what is expected for MCE.
Clearly, the MCE decreases with a decrease of the conductance peaks.
Otherwise, experiments on nanotubes with two ferromagnetic contacts
can be compared to those with one ferromagnet and a normal
metal.\cite{sahoonat} However, for a proper comparison, it is
essential, that the coupling to the normal metal and the ferromagnet
is very similar.
\\
Finally, we  discuss the influence of a demagnetizing field on the
MCE qualitatively. This field may play a significant role in
carbon nanotubes onto which a ferromagnetic strip is evaporated.
Locally, in the nanotube beneath the ferromagnet, the
demagnetizing field is expected to be quite high, of order 0.5 T
(assuming a field due to the ferromagnet of 1 T close to its
surface). The reason for this is that the aspect ratio of the
nanotube is unity (in the radial direction). The demagnetizing
field shifts the local work function of the ferromagnet thus
adding to MCE. Suppose now that the ferromagnet is magnetized in
the positive direction and a negative B field is applied. Then, we
expect the ferromagnetic domains in the vicinity of the nanotube
to change their orientation slowly. This locally rotates the
demagnetization field and therefore changes $\Delta q$. As a
consequence, a characteristic magnetoconductance trace is
expected, with conductance changes setting in \emph{before} the
ferromagnet actually switches (cf. Ref. \cite{brands}). As soon as
the ferromagnet does switch, we are in a mirror image of the
original situation and the contribution of the demagnetizing field
jumps back to its old value. We conclude that MCE due to the
demagnetizing field gives a continuous conductance change for
fields down to the coercive field. Just as for the
external-field-induced MCE, conductance changes are already
expected at fields close to 0 T. This is consistent with the
majority of two-terminal experiments.\cite{tsukagoshiapl,
tsukagoshinat,sahooapl,sahoonat,nagabhirava,yakushiji,zhang} In
Fig. \ref{vsB}b), we sketch the total MCE, including that of the
demagnitizing field (dashed line). We note the similarity of the
full MCE curve (though partly qualitative) with what is expected
for spin accumulation.\cite{voetsym}

In summary, we show that the magnetocoulomb effect should be taken
into account to explain experiments on spin valve structures in the
coulomb blockade regime. A proper separation of spin accumulation
and MCE is essential for a good understanding of the first.\\

\textit{Acknowledgements} This work was financed by the
Nederlandse Organisatie voor Wetenschappelijk Onderzoek, NWO, via
a Pionier grant.



\begin{thebibliography}{9}
\bibitem{voetP}
For MCE one should consider the thermodynamic quantity P, i.e., P
as a result of the full band structure. This P differs
considerably from the polarization determined in tunnel
experiments, since for the latter tunnel matrix elements play a
role as well. For Co, $P \approx -0.6$, whereas for Ni, $P \approx
-0.8$. See Ref. \cite{ono}.
\bibitem{voetCB}
We assume each coulomb peak to be independent. Although this is
not exactly correct, it suffices to show the principle of MCE.
\bibitem{voetq}
We use q as a relative quantity, i.e., $q=0$ corresponds to a situation where N electrons are present on the island (charge $-Ne$). N depends on the properties of the system and can be large. A change in q by e corresponds to a change in coulomb level energy by $e^2/C_{tot}$ (where $C_{tot}$ is the total capacitance of the island).
\bibitem{voetoffset}
We note that Fig. \ref{tmr}b) changes considerably for a sample
which exhibits a non-zero background conductance.
\bibitem{voetsym}
We choose a symmetric configuration. However, if asymmetric
contacts are assumed (e.g., $C_1 \neq C_2$), the shape of Fig.
\ref{tmr}b) will change accordingly. If one of the peaks in
Fig.\ref{tmr}b) dominates the other, only one peak may be observed
experimentally.
\end{thebibliography}
\end{document}